\documentclass{cvmp_short}
\usepackage{amssymb}
\usepackage{float}
\usepackage{graphicx}
\usepackage{svg}
\usepackage{graphicx}

\begin{document} \title{Image-based Facial Rig Inversion} 

\addauthor{
Tianxiang Yang \textsuperscript{1,2}\; 
Marco Volino \textsuperscript{1}\; 
Armin Mustafa \textsuperscript{1}\;\\
 Greg Maguire \textsuperscript{2}\; 
Robert Kosk{}}{}{2} 

\addinstitution{ University of Surrey, Guildford GU2 7XH, UK } \addinstitution{ Humain Ltd., Belfast BT1 4DD, UK }

\maketitle

\noindent\textbf{Introduction:}  
Facial rig inversion plays a key role in animation production, virtual avatars, and performance capture pipelines, where accurate recovery of control parameters of the rig from visual input enables direct control of production assets.  
Early approaches relied on statistical or regression models trained on animator-created data~\cite{Holden2015,Rackovic2021}.  
More recently, learning-based methods have enabled mesh-level supervision through differentiable rig approximations~\cite{Bolduc2022}.  
Most prior work relies on mesh-level features, which are information-rich but restricted to well-structured topology. In contrast, image-based inputs provide the advantage of generalizing to scanned data, a direction that remains underexplored.

\noindent\textbf{Method:}  
The image domain using two complementary renderings of a 3D facial mesh: a conventional RGB image appearance, and an RGB-encoded normal map describing per-pixel surface orientation. Each input is processed by an independent Hiera~\cite{Hiera2023} backbone, and the extracted features are concatenated and passed to a multi-layer perceptron (MLP) regression head to predict 102 rig control parameters derived from the Facial Action Coding System (FACS). As shown in Figure 1, the predicted parameters are decoded by a programmatic rig implemented in PyTorch, to reflect a custom Maya facial rig for mesh reconstruction. The model is supervised in rig parameters space and 3D mesh space, ensuring that the network learns values that are both numerically accurate and geometrically consistent.

\begingroup
\setlength{\intextsep}{3pt}
\begin{figure}[h]
\centering
\includegraphics[width=0.5\textwidth]{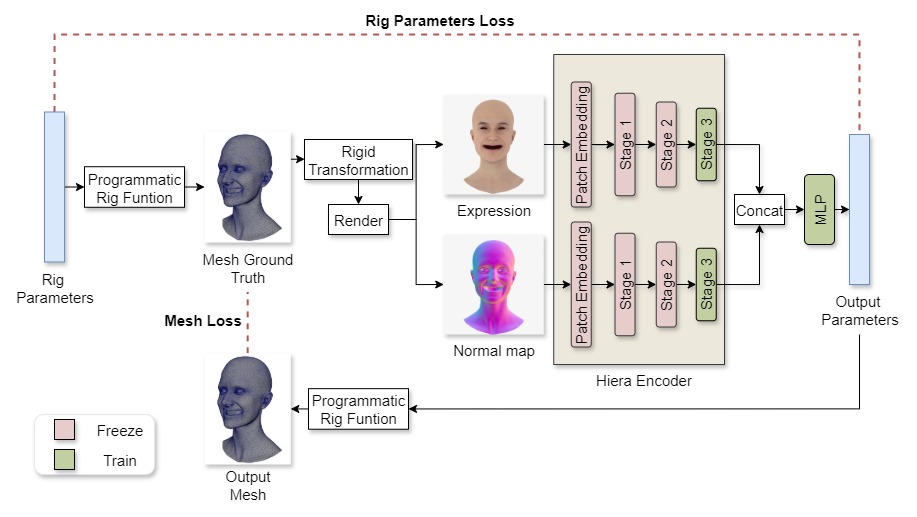}
\vspace{-20pt}
\caption{Proposed dual-branch network architecture. Each branch processes one image modality via a Hiera transformer; features are concatenated and regressed to rig parameters; and then reconstructed by programmatic rig function.}
\label{fig:network}
\end{figure}
\noindent\textbf{Dataset:} 
The dataset is divided into training and validation sets. The validation set contains scanned sequences of actors performing 20 expressions. The training set uses synthetic data from deformation transfer (DT)~\cite{sumner2004deformation} blendshapes rigs of the same characters. Each rig parameter is activated independently, along with 20 manual combinations representing canonical expressions. To mirror real performance variability, parameters are probabilistically dropped, added, or replaced. Parameter values are then sampled from a normal distribution to create different intensities. These sets are input into the rig function to reconstruct the mesh.
To improve robustness to subtle misalignment present in in-the-wild scanned data, the synthetic mesh is augmented with rigid transformations. 

Rendering is performed at a fixed resolution with a constant camera pose and consistent three-point lighting during both training and inference. Normal maps are encoded in the tangent space, mapping surface normals in $[-1, 1]$ to the $[0, 255]$ RGB range.

\noindent\textbf{Training:}  
Given the appearance image $I_a$ and the normal map $I_n$, the objective is to learn $f_\theta : (I_a, I_n) \rightarrow \mathbf{p} \in \mathbb{R}^{102}$.  
where $\mathbf{p}$ denotes the control parameters of the target rig.  All images are resized to $512\times512$ pixels, center-cropped and normalized using ImageNet statistics to match the backbone pretraining.  
Both RGB and normal map inputs are processed by independent Hiera~\cite{Hiera2023} backbones. Instead of the $224\times224$ resolution used in pre-training, we employ $512\times512$ inputs, as higher-resolution facial imagery preserves fine-grained texture and geometric cues that are critical to capture subtle expression changes. The patch embedding layer and the first three encoder stages in each backbone are frozen to retain the low-level appearance and geometric features, while the final stage remains trainable. 
\begin{figure}[t]
    \centering
    \includegraphics[width=0.95\linewidth]{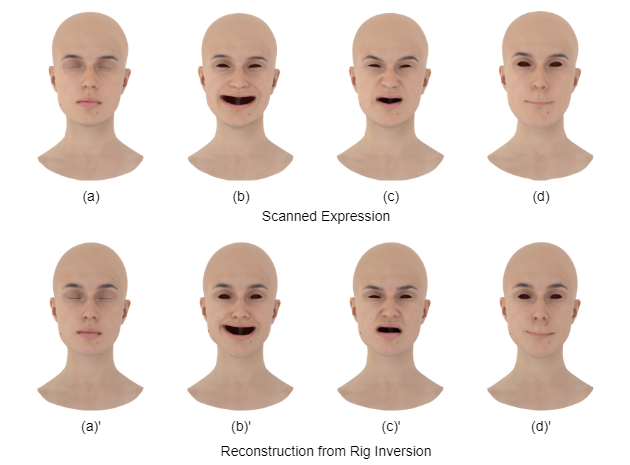}
     \vspace{-10pt}
    \caption{Input with scanned expression, reconstructed by proposed rig inversion method.}
     \vspace{-15pt}
    \label{fig:qual}
\end{figure}

During training, the network is supervised using mesh data and rig parameters. The overall objective combines a parameter-space loss and a mesh-space loss: a mean squared error (MSE) between the predicted and ground-truth rig parameters and a $L_{1}$ loss on the reconstructed mesh obtained through the programmatic rig. Optimization is performed using the AdamW optimizer with a learning rate of $1\times10^{-4}$. Training runs for 200 epochs with a batch size of 32. Training is executed on a single NVIDIA 4080 Laptop GPU. With 22575 training samples per branch, each epoch comprises 706 iterations, resulting in a total of approximately 141k training steps.

\noindent\textbf{Evaluation:}  
We evaluate the model on scanned data, and the predicted parameters are applied to the same DT blendshapes rig used in training to reconstruct the mesh. The reconstructed mesh is visualized by rendering, as shown in Figure~\ref{fig:qual}. The prediction is especially strong around mouth region. Gaze directions with upward, downward, or lateral motion are relatively more challenging for rig inversion.

\noindent\textbf{Conclusion:}  
This work introduces an image-based rig inversion framework that combines appearance and normal inputs to recover rig parameters. The approach generalizes to scanned data, producing faithful reconstructions. In future work, extending fine-tuning beyond the current partial-freeze strategy to the entire network may further improve adaptation to the rig inversion setting.
\vspace{10pt}

\noindent\textbf{References}  
\bibliography{egbib}

\end{document}